\documentclass[reqno,a4paper]{amsart}
\usepackage{amsmath}
\numberwithin{equation}{section}
\usepackage{amsfonts}
\usepackage{graphicx}
\usepackage{booktabs}
\usepackage{hyperref}
\hypersetup{
    colorlinks,%
    citecolor=blue,%
    filecolor=blue,%
    linkcolor=blue,%
    urlcolor=blue
}
\usepackage[authoryear]{natbib}
\usepackage{tikz}
\newcommand{\authorfootnotes}{\renewcommand\thefootnote{\@fnsymbol\c@footnote}}%

\newtheorem{proposition}{Proposition}[section]

\newtheorem{theorem}[proposition]{Theorem}

\newtheorem{remark}[proposition]{Remark}

\newcommand{\thlabel}[1]{\label{th:#1}}
\newcommand{\thref}[1]{Theorem~\ref{th:#1}}
\newcommand{\selabel}[1]{\label{se:#1}}
\newcommand{\seref}[1]{Section~\ref{se:#1}}

\newcommand{\relabel}[1]{\label{re:#1}}
\newcommand{\reref}[1]{Remark~\ref{re:#1}}

\newcommand{\eqlabel}[1]{\label{eq:#1}}
\newcommand{\equref}[1]{(\ref{eq:#1})}

\def\et{\eta}

\def\g{{\gamma}}
\def\a{{\alpha}}

\def\b{{\beta}}
\def\G{{\Gamma}}
\def\d{{\delta}}

\def\1{{\mathbf 1}}

\def\PP{{\mathbb P}}

\def\MR{{\mathcal R}}

\begin{document}
\bibliographystyle{plainnat}

\begin{center}
  \LARGE
{\sc Approximation for the Distribution of Three-dimensional Discrete Scan Statistic}\par \bigskip
  \normalsize
  Alexandru Am\u arioarei\textsuperscript{1,2,3}, Cristian Preda\textsuperscript{1,2}\par \bigskip

  \textsuperscript{1}Laboratoire de Math\'emathiques Paul Painlev\'e, UMR 8524, Universit\'e de Sciences et Technologies de Lille 1, France \par
  \textsuperscript{2}INRIA Nord Europe/Modal, France\par
  \textsuperscript{3}National Institute of R\&D for Biological Sciences, Bucharest, Romania\par \bigskip

  \today
\end{center}

\email{alexandru.amarioarei@inria.fr}
\email{cristian.preda@polytech-lille.fr}


\begin{abstract}
{We consider the discrete three dimensional scan statistics.
Viewed  as the maximum of an 1-dependent stationary r.v.'s sequence, we
provide approximations and error bounds for the probability distribution
of the three dimensional scan statistics. Importance sampling algorithm is used to obtains
sharp bounds for the simulation error. Simulation results and comparisons with other
approximations are presented for the binomial and Poisson models.}
\end{abstract}

\section{Introduction}\selabel{sec1}
\noindent
Let $T_1$, $T_2$, $T_3$ be positive integers, $\MR=[0,T_1]\times[0,T_2]\times[0,T_3]$ be a rectangular region and $\{X_{ijk}|1\leq i\leq T_1,1\leq j\leq T_2,1\leq k\leq T_3\}$ be a family of independent and identically distributed integer valued random variables from a specified distribution. In practice, $X_{ijk}$ can be interpreted as the number of events that occur in the elementary subregion $r_{ijk}=[i-1,i]\times[j-1,j]\times[k-1,k]$. For each $j\in\{1,2,3\}$, consider the positive integers $m_j$ such that $2\leq m_j\leq T_j-1$, and define the random variables
\begin{equation}\eqlabel{eq1}
Y_{i_1i_2i_3}=\displaystyle\sum_{i=i_1}^{i_1+m_1-1}{\sum_{j=i_2}^{i_2+m_2-1}{\sum_{k=i_3}^{i_3+m_3-1}{X_{ijk}}}},\ \ \ \ 1\leq i_j\leq T_j-m_j+1,
\end{equation}
as the number of events occurring in the rectangular region
$$
\MR(i_1,i_2,i_3)=[i_1-1,i_1+m_1-1]\times[i_2-1,i_2+m_2-1]\times[i_3-1,i_3+m_3-1].
$$
The three dimensional discrete scan statistic is defined as the maximum number of events in any rectangle $\MR(i_1,i_2,i_3)$ within the region $\MR$,
\begin{equation}\eqlabel{eq2}
S_{m_1,m_2,m_3}(T_1,T_2,T_3)=\max_{\substack{1\leq i_j\leq T_j-m_j+1\\j\in\{1,2,3\}}}{Y_{i_1i_2i_3}}.
\end{equation}
The distribution of scan statistics,
\begin{equation*}
\displaystyle\PP\left(S_{m_1,m_2,m_3}(T_1,T_2,T_3)\leq n\right),\ \ \ n\in\{1,2,\dots,m_1m_2m_3\}
\end{equation*}
is used with success in astronomy (\citet{Darling1986}), image analysis and reliability theory (\citet{Boutsikas2000}) and many other domains. For an overview of the potential application of scan statistics one can refer to the monographs of \citet{Glaz2001} and more recently the one of \citet{Glaz2009}.
\par\noindent
From a statistical point of view, the scan statistic $S_{m_1,m_2,m_3}(T_1,T_2,T_3)$ is used for testing the null hypothesis of randomness that $X_{ijk}$'s are independent and identically distributed according to some specified distribution. Under the alternative hypothesis there exists one cluster location where the $X_{ijk}$'s have a larger mean than outside the cluster. As an example, in the Poisson model, the null hypothesis, $H_0$, assumes that $X_{ijk}$'s are i.i.d. with $X_{ijk}\sim Pois(\lambda)$ whereas the alternative hypothesis of clustering, $H_1$, assumes the existence of a rectangular subregion $\MR(i_0,j_0,k_0)$ such that for any $i_0\leq i\leq i_0+m_1-1$, $j_0\leq j\leq j_0+m_2-1$ and $k_0\leq k\leq k_0+m_3-1$, $X_{ijk}$ are i.i.d. Poisson random variables with parameter $\lambda'>\lambda$. Outside the region $\MR(i_0,j_0,k_0)$, $X_{ijk}$ are i.i.d. distributed according to the distribution specified by the null hypothesis. The generalized likelihood ratio test rejects $H_0$ in favor of the local change alternative $H_1$, whenever $S_{m_1,m_2,m_3}(T_1,T_2,T_3)$ exceeds the threshold $\tau$ determined from $\PP\left(S_{m_1,m_2,m_3}(T_1,T_2,T_3)\geq \tau|H_0\right)=\alpha$ and where $\alpha$ represents the significance level of the testing procedure (\citet[Chapter 13]{Glaz2001}).
\par\noindent
Since there are no exact formulas available for the distribution of three dimensional scan statistics, approximation methods are necessary. For the Bernoulli model, \citet{Glaz2010} propose four approximation formulas: one Markov like product type approximation and three Poisson type approximations that extends the special case described by \citet{Darling1986} when $n=m_1m_2m_3$.
\par\noindent
The advantage of the method described in this paper is that it can be used for any distribution of the random field and provides accurate approximations and sharp error bounds. The methodology used to obtain the approximation and the error bounds is presented in \seref{sec2}. In \seref{sec3} we describe adapt the importance sampling algorithm developed by \citet{Naiman2001} to estimate the simulation error. A simulation study is conducted in \seref{sec4} for considered Bernoulli, binomial and Poisson models. Concluding remarks are given in \seref{sec5}.
\section{Methodology}\selabel{sec2}
\noindent
In order to approximate the distribution of $S_{m_1,m_2,m_3}(T_1,T_2,T_3)$ we use a similar approach as in \citet{HaimanPreda2}. The key idea is to observe that we can write the scan statistic random variable as a maximum of 1-dependent stationary sequence of random variables. A sequence $(Z_k)_{k\geq1}$ is $m$-dependent, $m\geq1$, if for any $h\geq1$ the $\sigma$-fields generated by $\{Z_1,\dots,Z_h\}$ and $\{Z_{h+m+1},\dots\}$ are independent. The method is based on the following result developed in \citet[Theorem 4]{Haiman} and improved in \citet[Theorem 2.6]{Amarioarei}:\\
Let $(Z_k)_{k\geq1}$ be a strictly stationary 1-dependent sequence of random variables and for $x<\sup\{u|\PP(Z_1\leq u)<1\}$, let
\begin{equation}\eqlabel{meq1}
q_m=q_m(x)=\PP(\max(Z_1,\dots,Z_m)\leq x).
\end{equation}

\begin{figure}[ht]
\begin{tikzpicture}[scale=0.6,every node/.style={scale=0.6}]

\begin{scope}[color=gray!60]
   \begin{scope}[rotate=0,black,inner sep=2pt]
        \draw[dashed, black!40] (1.155,-0.577) -- +(-0.7,-0.4)
            node [coordinate, near end] (a) {};
        \draw[dashed, black!40] (8.9,-4.4) -- +(-0.7,-0.4)
            node [coordinate, near end] (b) {};
        \draw[|<->|] (a) -- node[fill=white] {\large$T_2$} (b);
    \end{scope}

    \begin{scope}[rotate=0,black,inner sep=2pt]
        \draw[dashed, black!40] (8.9,-4.4) -- +(0.8,-0.4)
            node [coordinate, near end] (a) {};
        \draw[dashed, black!40] (16.9,-0.46) -- +(0.8,-0.4)
            node [coordinate, near end] (b) {};
        \draw[|<->|] (a) -- node[fill=white] {\large$T_1$} (b);
    \end{scope}

    \begin{scope}[rotate=0,black,inner sep=2pt]
        \draw[dashed, black!40] (16.9,-0.46) -- +(0.5,0)
            node [coordinate, near end] (a) {};
        \draw[dashed, black!40] (16.9,7.56) -- +(0.5,0)
            node [coordinate, near end] (b) {};
        \draw[|<->|] (a) -- node[fill=white] {\large$T_3$} (b);
    \end{scope}
\begin{scope}
		\draw[fill,black] (1.15,-0.55) circle [radius=0.07];
		\node [left,black] at (1.15,-0.55) {\normalsize$1$};
		
		\draw[fill,black] (1.15,1.05) circle [radius=0.07];
		\node [left,black] at (1.15,0.95) {\normalsize$2(m_3-1)$};
		
		\draw[fill,black] (1.15,0.42) circle [radius=0.07];
		\node [left,black] at (1.15,0.42) {\normalsize$m_3$};
		
		\draw[fill,black] (1.15,2.05) circle [radius=0.07];
		\node [left,black] at (1.15,2.05) {\normalsize$3(m_3-1)$};
		
		\draw[fill,black] (1.15,1.15) circle [radius=0.07];
		\node [left,black] at (1.15,1.25) {\normalsize$2m_3-1$};
		
		\draw[fill,black] (1.15,2.7) circle [radius=0.07];
		\node [left,black] at (1.15,2.7) {\normalsize$4(m_3-1)$};
		
    	\end{scope}

\begin{scope}[yslant=-0.5,xscale=0.77,shift={(1.5,0)}]
  \draw[fill=blue!10](0,0) rectangle +(10,1.6);
\end{scope}

\begin{scope}[yslant=0.5,shift={(0.7,-0.7)}]
\draw[fill=blue!10](8.15,-8.15) rectangle +(8,1.6);
\end{scope}

\begin{scope}[yslant=0.5,xslant=-1,yscale=0.77,shift={(-0.4,-1.17)}]
\draw[fill=blue!10](10,1.75) rectangle +(-8,-10);
\end{scope}
    \begin{scope}[yslant=-0.5,xscale=0.77,shift={(1.5,0)}]
                 \draw[fill=green!40,opacity=0.3](0,1) rectangle +(10,1.6);
    \end{scope}

    \begin{scope}[yslant=0.5,shift={(0.7,-0.7)}]
        \draw[fill=green!40,opacity=0.3] (8.15,-7.15) rectangle +(8,1.6);
    \end{scope}

    \begin{scope}[yslant=0.5,xslant=-1,yscale=0.77,shift={(-0.38,-1.4)}]
        \draw[fill=green!40,opacity=0.3] (11,3.3) rectangle +(-8,-10);
    \end{scope}

    \begin{scope}[yslant=-0.5,xscale=0.77,shift={(1.5,0.1)}]
                 \draw[fill=red!30,opacity=0.3](0,1.6) rectangle +(10,1.6);
    \end{scope}

    \begin{scope}[yslant=0.5,shift={(0.7,-0.6)}]
        \draw[fill=red!30,opacity=0.3] (8.15,-6.55) rectangle +(8,1.6);
    \end{scope}

    \begin{scope}[yslant=0.5,xslant=-1,yscale=0.77,shift={(-0.8,-1.85)}]
        \draw[fill=red!30,opacity=0.3] (12.1,4.65) rectangle +(-8,-10);
    \end{scope}

	\begin{scope}[yslant=-0.5,xscale=0.77,shift={(1.5,0)}]
 	 	\draw[thick,gray] (0,0) grid (10,8);
	\end{scope}

	\begin{scope}[yslant=0.5,shift={(-1.15,-0.85)}]
  		\draw[thick,gray] (10,-8) grid (18,0);
	\end{scope}

	\begin{scope}[yslant=0.5,xslant=-1,yscale=0.77,shift={(-2,-1.1)}]
  		\draw[thick,gray] (10,0) grid (18,10);
	\end{scope}

	\begin{scope}[yslant=0.5,shift={(1.15,-1.15)}]
 		 \draw[dashed,gray] (0,0) grid (8,8);
	\end{scope}

	\begin{scope}[yslant=0.5,xslant=-1,yscale=0.77,shift={(0,-1.5)}]
  		\draw[dashed,gray] (0,-10) grid (8,0);
	\end{scope}

	\begin{scope}[yslant=-0.5,xscale=0.77,shift={(1.88,-2)}]
 		 \draw[dashed,gray] (10,10) grid (20,18);
	\end{scope}

    \begin{scope}[rotate=0,black,inner sep=2pt]
        \draw[dashed, black!20] (16.9,1.2) -- +(1.8,0)
            node [coordinate, near end] (a) {};
        \draw[dashed, black!20] (16.9,2.9) -- +(1.8,0)
            node [coordinate, near end] (b) {};
        \draw[|<->|] (a) -- node[fill=white] {\normalsize$Z_3$} (b);
    \end{scope}

    \begin{scope}[rotate=0,black,inner sep=2pt]
        \draw[dashed, black!20] (16.9,0.55) -- +(1.2,0)
            node [coordinate, near end] (a) {};
        \draw[dashed, black!20] (16.9,2.15) -- +(1.2,0)
            node [coordinate, near end] (b) {};
        \draw[|<->|] (a) -- node[fill=white] {\normalsize$Z_2$} (b);
    \end{scope}

     \begin{scope}[rotate=0,black,inner sep=2pt]
        \draw[dashed, black!20] (16.9,-0.45) -- +(0.8,0)
            node [coordinate, near end] (a) {};
        \draw[dashed, black!20] (16.9,1.175) -- +(0.8,0)
            node [coordinate, near end] (b) {};
        \draw[|<->|] (a) -- node[fill=white] {\normalsize$Z_1$} (b);
    \end{scope}

\begin{scope}
    \begin{scope}[yslant=-0.5,xscale=0.77,shift={(1.5,0)}]
         \draw[blue!50!black,very thick,dashed](0,0) rectangle +(10,1.6);
    \end{scope}

    \begin{scope}[yslant=0.5,shift={(0.7,-0.7)}]
        \draw[blue!50!black,very thick,dashed](8.15,-8.15) rectangle +(8,1.6);
    \end{scope}

    \begin{scope}[yslant=0.5,xslant=-1,yscale=0.77,shift={(-0.4,-1.17)}]
        \draw[blue!50!black,very thick,dashed,opacity=0.7](10,1.75) rectangle +(-8,-10);
    \end{scope}
 \end{scope}
    \begin{scope}[yslant=-0.5,xscale=0.77,shift={(1.5,0)}]
         \draw[green!20!black,very thick,dashed](0,1) rectangle +(10,1.6);
    \end{scope}

    \begin{scope}[yslant=0.5,shift={(0.7,-0.7)}]
        \draw[green!20!black,very thick,dashed](8.15,-7.15) rectangle +(8,1.6);
    \end{scope}

    \begin{scope}[yslant=0.5,xslant=-1,yscale=0.77,shift={(-0.38,-1.4)}]
       \draw[green!20!black,ultra thick,dashed,opacity=0.7] (11,3.3) rectangle +(-8,-10);
    \end{scope}
     \begin{scope}[yslant=-0.5,xscale=0.77,shift={(1.5,0.1)}]
          \draw[red!60!black,ultra thick,dashed](0,1.6) rectangle +(10,1.6);
     \end{scope}

     \begin{scope}[yslant=0.5,shift={(0.7,-0.6)}]
         \draw[red!60!black,very thick,dashed](8.15,-6.55) rectangle +(8,1.6);
     \end{scope}

     \begin{scope}[yslant=0.5,xslant=-1,yscale=0.77,shift={(-0.8,-1.85)}]
        \draw[red!60!black,ultra thick,dashed,opacity=0.7] (12.1,4.65) rectangle +(-8,-10);
     \end{scope}
\begin{scope}[shift={(-1.5,10)},scale=1.4]

\begin{scope}[rotate=0,black,inner sep=1pt]
        \draw[dashed, black!20] (1.95,-0.95) -- +(0.7,-0.3)
            node [coordinate, near end] (a) {};
        \draw[dashed, black!20] (2.95,-0.45) -- +(0.7,-0.3)
            node [coordinate, near end] (b) {};
        \draw[|<->|] (a) -- node[fill=white] {\normalsize$m_1$} (b);
    \end{scope}

    \begin{scope}[rotate=0,black,inner sep=1pt]
        \draw[dashed, black!20] (1.155,-0.577) -- +(-0.7,-0.3)
            node [coordinate, near end] (a) {};
        \draw[dashed, black!20] (1.95,-0.95) -- +(-0.7,-0.3)
            node [coordinate, near end] (b) {};
        \draw[|<->|] (a) -- node[fill=white] {\normalsize$m_2$} (b);
    \end{scope}

     \begin{scope}[rotate=0,black,inner sep=1pt]
        \draw[dashed, black!20] (1.155,-0.577) -- +(-0.7,0)
            node [coordinate, near end] (a) {};
        \draw[dashed, black!20] (1.155,0.42) -- +(-0.7,0)
            node [coordinate, near end] (b) {};
        \draw[|<->|] (a) -- node[fill=white] {\normalsize$m_3$} (b);
    \end{scope}

\begin{scope}[yslant=-0.5,xscale=0.77,shift={(1.5,0)}]
  \draw[thick,gray](0,0) rectangle +(1,1);
\end{scope}

\begin{scope}[yslant=0.5,shift={(0.7,-0.7)}]
\draw[thick,gray](1.22,-1.22) rectangle +(1,1);
\end{scope}

\begin{scope}[yslant=0.5,xslant=-1,yscale=0.77,shift={(0,-0.9)}]
\draw[thick,gray](2,0.7) rectangle +(-1,-1);
\end{scope}

\end{scope}
\end{scope}
\end{tikzpicture}
  \caption{Illustration of $Z_k$ emphasizing the $1$-dependence}
  \label{fig1}
\end{figure}
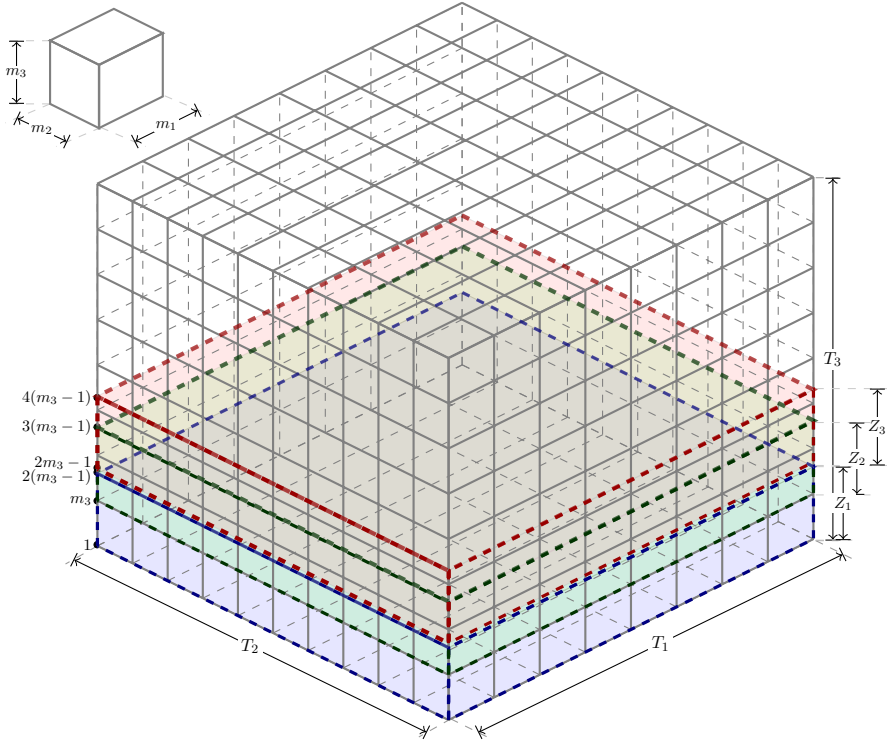

\begin{theorem}\thlabel{T1}
For all $x$ such that $q_1(x)\geq1-\a\geq 0.9$, the following approximation formula holds:
\begin{equation}\eqlabel{eqT1}
\left|q_m-\frac{2q_1-q_2}{\left[1+q_1-q_2+2(q_1-q_2)^2\right]^m}\right|\leq mF(\a,m)(1-q_1)^2
\end{equation}
with
\begin{equation}\eqlabel{eqT1.2}
F(\a,m)=1+\frac{3}{m}+\left[\frac{\G(\a)}{m}+K(\a)\right](1-q_1)
\end{equation}
where $\G(\a)=L(\a)+E(\a)$,
\begin{align}
K(\a)&=\frac{\frac{11-3\a}{(1-\a)^2}+2l(1+3\a)\frac{2+3l\a-\a(2-l\a)(1+l\a)^2}{\left[1-\a(1+l\a)^2\right]^3}}{1-\frac{2\a(1+l\a)}{\left[1-\a(1+l\a)^2\right]^2}}\eqlabel{eqK1}\\
L(\a)&=3K(\a)(1+\a+3\a^2)[1+\a+3\a^2+K(\a)\a^3]+\a^6K^3(\a)\nonumber\\
     & +9\a(4+3\a+3\a^2)+55.1 \eqlabel{eqL1}\\
E(\a)&=\frac{\et^5\left[1+(1-2\a)\et\right]^4\left[1+\a(\et-2)\right]\left[1+\et+(1-3\a)\et^2\right]}{2(1-\a\et^2)^4\left[(1-\a\et^2)^2-\a\et^2(1+\et-2\a\et)^2\right]}\eqlabel{eqE1}
\end{align}
and where $\et=1+l\a$ with $l=l(\a)>t_2^3(\a)$ and $t_2(\a)$ the second root in magnitude of the equation $\a t^3-t+1=0$.
\end{theorem}
\noindent
In this section we obtain an approximation formula for the distribution of scan statistic defined by Eq.\equref{eq2} in three steps as follows.\\
Let assume that $L_j=\frac{T_j}{m_j-1}$, $j\in\{1,2,3\}$, are positive integers and define for each $k\in\{1,2,\dots,L_3-1\}$ the random variables
\begin{equation}\eqlabel{meq2}
Z_k=\max_{\substack{1\leq i_1\leq(L_1-1)(m_1-1)\\1\leq i_2\leq(L_2-1)(m_2-1)\\(k-1)(m_3-1)+1\leq i_3\leq k(m_3-1)}}{Y_{i_1i_2i_3}}.
\end{equation}
 The set of random variables $\{Z_1,\dots,Z_{L_3-1}\}$ forms a 1-dependent stationary sequence. Indeed, from Eq.\equref{meq2} and the independence of $X_{ijl}$ we observe that for any $k\geq1$, $\sigma(\cdots,Z_k)$ and $\sigma(Z_{k+2},\cdots)$ are included in $\sigma(\{X_{ijl}|1\leq i\leq T_1,1\leq j\leq T_2,1\leq l\leq (k+1)(m_3-1)\})$ and $\sigma(\{X_{ijl}|1\leq i\leq T_1,1\leq j\leq T_2,(k+1)(m_3-1)+1\leq l\})$, respectively, which are independent (see Fig.~\ref{fig1}).
\par\noindent
Notice that from Eq.\equref{eq2} and Eq.\equref{meq2} we have
\begin{equation}\eqlabel{meq3}
S(L_1,L_2,L_3)=S_{m_1,m_2,m_3}(T_1,T_2,T_3)=\max_{1\leq k\leq L_3-1}{Z_k}.
\end{equation}
Take for $s\in\{2,3\}$,
\begin{equation}\eqlabel{meq4}
Q_s=Q_s(n)=\displaystyle\PP\left(\bigcap_{k=1}^{s-1}\{Z_k\leq n\}\right)=\displaystyle\PP\left(\max_{\substack{1\leq i_1\leq(L_1-1)(m_1-1)\\1\leq i_2\leq(L_2-1)(m_2-1)\\1\leq i_3\leq (s-1)(m_3-1)}}{Y_{i_1i_2i_3}}\leq n\right).
\end{equation}
Notice that in the notation of Eq.\equref{meq1} we have $Q_s=q_{s-1}$. For $n$ such that $Q_2(n)\geq1-\a_1\geq0.9$ we apply the result in \thref{T1} to obtain the first step approximation
\begin{equation}\eqlabel{meq5}
\PP\left(S(L_1,L_2,L_3)\leq n\right)\approx \frac{2Q_2-Q_3}{\left[1+Q_2-Q_3+2(Q_2-Q_3)^2\right]^{(L_3-1)}},
\end{equation}
with an error bound of $(L_3-1)F(\a_1,L_3-1)(1-Q_2)^2$. Observe that $Q_2$ and $Q_3$ represents the distribution of the scan statistics over the rectangular subregions $\left[1,T_1\right]\times\left[1,T_2\right]\times\left[1,2(m_3-1)\right]$ and $\left[1,T_1\right]\times\left[1,T_2\right]\times\left[1,3(m_3-1)\right]$, respectively (see also Fig~\ref{fig1}). To simplify the results of the presentation, in what follows we abbreviate the approximation formula by
\begin{equation}
H(x,y,m)=\frac{2x-y}{[1+x-y+2(x-y)^2]^{m-1}}.
\end{equation}
\noindent
In order to evaluate the approximation in Eq.\equref{meq5} it is necessary to find approximations for $Q_2$ and $Q_3$. Thus, the second step consists in applying \thref{T1} for each $Q_s$. We define, as in Eq.\equref{meq2}, for $s\in\{2,3\}$ and $l\in\{1,2,\dots,L_2-1\}$ the sequences
\begin{equation}\eqlabel{meq6}
Z^{(s)}_{l}=\max_{\substack{1\leq i_1\leq(L_1-1)(m_1-1)\\(l-1)(m_2-1)+1\leq i_2\leq l(m_2-1)\\1\leq i_3\leq (s-1)(m_3-1)}}{Y_{i_1i_2i_3}},
\end{equation}
which are strictly stationary, 1-dependent and satisfy
\begin{equation}\eqlabel{meq7}
Q_s=\PP(S(L_1,L_2,s)\leq n)=\PP\left(\max_{1\leq l\leq L_2-1}{Z^s_l}\leq n\right).
\end{equation}
Set for $t,s\in\{2,3\}$,
\begin{equation}\eqlabel{meq8}
Q_{ts}=Q_{ts}(n)=\PP\left(\bigcap_{l=1}^{t-1}\{Z^{(s)}_l\leq n\}\right)=\displaystyle\PP\left(\max_{\substack{1\leq i_1\leq(L_1-1)(m_1-1)\\1\leq i_2\leq (t-1)(m_2-1)\\1\leq i_3\leq (s-1)(m_3-1)}}{Y_{i_1i_2i_3}}\leq n\right).
\end{equation}
If the condition $Q_{2s}(n)\geq1-\a_{2s}\geq0.9$ is fulfilled, then using \thref{T1}, we find, for $s\in\{2,3\}$, the approximation
\begin{equation}\eqlabel{meq9}
\left|Q_s-H\left(Q_{2s},Q_{3s},L_2\right)\right|\leq (L_2-1)F(\a_{2s},L_2-1)(1-Q_{2s})^2.
\end{equation}
The last step involves the evaluation of $Q_{ts}$ in Eq.\equref{meq9}. For $s,t\in\{2,3\}$ and $j\in\{1,2,\dots,L_1-1\}$ let consider the following random sequences
\begin{equation}\eqlabel{meq10}
Z^{(ts)}_{j}=\max_{\substack{(j-1)(m_1-1)+1\leq i_1\leq j(m_1-1)\\1\leq i_2\leq (t-1)(m_2-1)\\1\leq i_3\leq (s-1)(m_3-1)}}{Y_{i_1i_2i_3}}.
\end{equation}
We observe that $\left(Z^{(ts)}_{j}\right)_{j\geq1}$ forms $1$-dependent stationary sequences and
\begin{equation}\eqlabel{meq11}
Q_{ts}=\PP\left(S(L_1,t,s)\leq n\right)=\PP\left(\max_{1\leq j\leq L_1-1}{Z^{(ts)}_j}\leq n\right).
\end{equation}
Put for $r,t,s\in\{2,3\}$
\begin{equation}\eqlabel{meq12}
Q_{rts}=Q_{rts}(n)=\displaystyle\PP\left(\bigcap_{j=1}^{r-1}\{Z^{ts}_j\leq n\}\right)=\displaystyle\PP\left(\max_{\substack{1\leq i_1\leq (r-1)(m_1-1)\\1\leq. i_2\leq(t-1)(m_2-1)\\1\leq i_3\leq(s-1)(m_3-1)}}{Y_{i_1i_2i_3}}\leq n\right)
\end{equation}
Then, under supplementary condition that $Q_{2ts}\geq1-\a_{3ts}\geq0.9$, we apply the result in \thref{T1} to obtain
\begin{equation}\eqlabel{meq13}
\left|Q_{ts}-H\left(Q_{2ts},Q_{3ts},L_1\right)\right|\leq (L_1-1)F(\a_{3ts},L_1-1)(1-Q_{2ts})^2.
\end{equation}
Combining the Eqs.\equref{meq5}, \equref{meq9} and \equref{meq13} we obtain an approximation formula for the distribution of the scan statistic depending on the eight quantities $Q_{rts}$, that we propose to evaluated by simulation. Note that in the above approximations, at each step we consider different values for $\a$. In the next section we show how to choose these values.
\begin{remark}\relabel{rem1}
If $T_1$, $T_2$ and $T_3$ are not multiples of $m_1-1$, $m_2-1$ and $m_3-1$, respectively, then let consider $L_j=\left\lfloor{\frac{T_j}{m_j-1}}\right\rfloor$ for $j\in\{1,2,3\}$. Based on the inequalities
\begin{equation}\eqlabel{meq14}
\PP(S(L_1+1,L_2+1,L_3+1)\leq n)\leq\PP\leq\PP(S(L_1,L_2,L_3)\leq n),
\end{equation}
we can approximate $\PP=\PP(S_{m_1,m_2,m_3}(T_1,T_2,T_3)\leq n)$ by linear interpolation (see Table~\ref{tab:Tb3}).
\end{remark}
\subsection{Computing the approximation error}\selabel{subsec21}
\noindent
To simplify the presentation and the derivation of the approximation formulae, it is convenient to introduce the following notations for $s,t\in\{2,3\}$:
\begin{align}
\a_3&=1-Q_3,\ \a_{23}=1-Q_{23},\ \a_{233}=1-Q_{233},\nonumber\\
\g_{ts}&=H(Q_{2ts},Q_{3ts},L_1),\ \g_s=H(\g_{2s},\g_{3s},L_2),\nonumber\\
F_1  &=F(\a_3,L_3-1),\ F_2=F(\a_{23},L_2-1),\ F_3=F(\a_{233},L_1-1).\nonumber
\end{align}
It is not hard to see that $Q_3\leq Q_2$, $Q_{23}\leq Q_{22}$ and $Q_{233}\leq Q_{2ts}$, so that the choice for the thresholds $\a_3$, $\a_{23}$ and $\a_{233}$ becomes natural. Based on the mean value theorem in two dimensions, one can easily verify that for $m\geq 6$ and $y_i\leq x_i$, $i\in\{1,2\}$ we have the inequality:
\begin{equation}\eqlabel{pf2}
\left|H(x_1,y_1,m)-H(x_2,y_2,m)\right|\leq (m-2)\left[|x_1-x_2|+|y_1-y_2|\right].
\end{equation}
\noindent
In what follows we use the result from Eq.\equref{pf2} without restrictions. This is in agreement with the numerical values considered in \seref{sec4}.
We begin by observing that applying Eq.\equref{pf2} into Eq.\equref{meq5} we obtain
\begin{align}\eqlabel{cae1}
\left|\PP-H\left(\g_2,\g_3,L_3\right)\right|&\leq\left|\PP-H\left(Q_2,Q_3,L_3\right)\right|+\left|H\left(Q_2,Q_3,L_3\right)-H\left(\g_2,\g_3,L_3\right)\right|\nonumber\\
                                              &\leq(L_3-1)F_1\left(1-Q_2\right)^2+(L_3-2)\left[|Q_2-\g_2|+|Q_3-\g_3|\right],
\end{align}
where for simplicity we used the notation $\PP=\PP\left(S(L_1,L_2,L_3)\leq n\right)$. In the same manner, one can see that for $s\in\{2,3\}$ we have
\begin{align}\eqlabel{cae2}
\left|Q_s-\g_s\right|&\leq\left|Q_s-H\left(Q_{2s},Q_{3s},L_2\right)\right|+\left|H\left(Q_{2s},Q_{3s},L_2\right)-H\left(\g_{2s},\g_{3s},L_2\right)\right|\nonumber\\
                     &\leq(L_2-1)F_2\left(1-Q_{2s}\right)^2+(L_2-2)\left[|Q_{2s}-\g_{2s}|+|Q_{3s}-\g_{3s}|\right].
\end{align}
We notice that Eq.\equref{meq13} can be rewritten as
\begin{equation}\eqlabel{cae3}
\left|Q_{ts}-\g_{ts}\right|\leq (L_1-1)F_3(1-Q_{2ts})^2,\ \ s,t\in\{2,3\}.
\end{equation}
Finally, in order to find the approximation error it is sufficient to determine bounds for $1-Q_2$ and $1-Q_{2s}$. It can be easily checked that
\begin{equation}\eqlabel{cae4}
1-Q_{2s}\leq 1-\g_{2s}+|Q_{2s}-\g_{2s}|\leq \d_{2s}
\end{equation}
where
\begin{equation}\eqlabel{cae5}
\d_{2s}=1-\g_{2s}+(L_1-1)F_3(1-Q_{22s})^2.
\end{equation}
Similarly, we can write
\begin{equation}\eqlabel{cae6}
1-Q_{2}\leq 1-\g_{2}+|Q_{2}-\g_{2}|\leq \d_{2},
\end{equation}
with
\begin{equation}\eqlabel{cae7}
\d_2=1-\g_2+(L_2-1)F_2\d_{22}+(L_2-2)(L_1-1)F_3\left[(1-Q_{222})^2+(1-Q_{232})^2\right].
\end{equation}
Substituting Eqs.\equref{cae2}, \equref{cae3}, \equref{cae4} and \equref{cae6} in Eq.\equref{cae1} we derive the formula for the approximation error
\begin{align}\eqlabel{cae8}
 E_{app}&=(L_3-1)F_1\d_{2}^2+(L_3-2)(L_2-1)F_2\left(\d_{22}^2+\d_{23}^2\right)+\nonumber\\
 &+(L_3-2)(L_2-2)(L_1-1)F_3\left[\displaystyle\sum_{t,s\in\{2,3\}}{(1-Q_{2ts})^2}\right].
 \end{align}
\subsection{Computing the simulation errors}\selabel{subsec22}
Since, from our knowledge, there are no exact formulas available for the computation of $Q_{rts}$ we propose to evaluate them by simulation. It is obvious that the simulation error appears from two terms: first, from the approximation formula in Eq.\equref{cae1} and second, from the error bound in Eq.\equref{cae8}.
\par\noindent
Usually, between the true and the estimated value we have a relation of the form
\begin{equation}\eqlabel{cse1}
\left|Q_{rts}-\hat{Q}_{rts}\right|\leq \b_{rts},\ \ \ r,t,s\in\{2,3\}
\end{equation}
where $\hat{Q}_{rts}$ are the simulated values corresponding to $Q_{rts}$. Provided a simulation error bound $\b_{rts}$ as in Eq.\equref{cse1}, let denote the simulated values by
\begin{align*}
\hat{Q}_{ts}&=H(\hat{Q}_{2ts},\hat{Q}_{3ts},L_1),\\
\hat{Q}_s&=H(\hat{Q}_{2s},\hat{Q}_{3s},L_2).
\end{align*}
From Eq.\equref{pf2} one obtains
\begin{equation}\eqlabel{cse2}
\left|H\left(\g_2,\g_3,L_3\right)-H\left(\hat{Q}_2,\hat{Q}_3,L_3\right)\right|\leq(L_3-2)\left[\left|\g_2-\hat{Q}_2\right|+\left|\g_3-\hat{Q}_3\right|\right].
\end{equation}
Observe that the differences in the right hand term in Eq.\equref{cse2} can be bounded by
\begin{align}\eqlabel{cse3}
\left|\g_s-\hat{Q}_s\right|&=\left|H\left(\g_{2s},\g_{3s},L_2\right)-H\left(\hat{Q}_{2s},\hat{Q}_{3s},L_2\right)\right|\nonumber\\
                           &\leq(L_2-2)\left[\left|\g_{2s}-\hat{Q}_{2s}\right|+\left|\g_{3s}-\hat{Q}_{3s}\right|\right].
\end{align}
In the same way we can write for $t,s\in\{2,3\}$
\begin{align}\eqlabel{cse4}
\left|\g_{ts}-\hat{Q}_{ts}\right|&=\left|H\left(\g_{2ts},\g_{3ts},L_1\right)-H\left(\hat{Q}_{2ts},\hat{Q}_{3ts},L_1\right)\right|\nonumber\\
                           &\leq(L_1-2)\left[\left|\g_{2ts}-\hat{Q}_{2ts}\right|+\left|\g_{3ts}-\hat{Q}_{3ts}\right|\right]\nonumber\\
                           &\leq(L_1-2)\left[\b_{2ts}+\b_{3ts}\right].
\end{align}
Combining Eqs.\equref{cse4}, \equref{cse3} and \equref{cse2} we get the simulation error corresponding to the approximation formula
\begin{equation}\eqlabel{cse5}
E_{sf}=(L_1-2)(L_2-2)(L_3-2)\left(\displaystyle\sum_{r,t,s\in\{2,3\}}{\b_{rts}}\right).
\end{equation}
In order to obtain the simulation error corresponding to the approximation error bound in Eq.\equref{cae8} we follow the lines of \seref{subsec21}. With the following notations
\begin{align*}
u_{rts}&=1-\hat{q}_{rts}+\b_{rts},\\
u_{ts}&=1-\hat{q}_{ts}+(L_1-2)(\b_{2ts}+\b_{3ts}),\\
u_s&=1-\hat{q}_s+(L_1-2)(L_2-2)(\b_{22s}+\b_{32s}+\b_{23s}+\b_{33s}),\\
\bar{\d}_{2s} &= u_{2s}+(L_1-1)F_3u_{22s}^2,\\
\bar{\d}_{2} &= u_2+(L_2-1)F_2\bar{\d}_{22}+(L_2-2)(L_1-1)F_3(u_{222}^2+u_{232}^2),
\end{align*}
the error can be expressed as
\begin{align}\eqlabel{cse6}
 E_{sapp}&=(L_3-1)F_1\bar{\d}_{2}^2+(L_3-2)(L_2-1)F_2\left(\bar{\d}_{22}^2+\bar{\d}_{23}^2\right)+\nonumber\\
                            &+(L_3-2)(L_2-2)(L_1-1)F_3\left(\displaystyle\sum_{t,s\in\{2,3\}}{u_{2ts}^2}\right).
\end{align}
The total simulation error is obtained by adding the two terms from Eq.\equref{cse5} and Eq.\equref{cse6}
\begin{equation}\eqlabel{cse7}
E_{sim}=E_{sf}+E_{sapp}.
\end{equation}
To evaluate Eq.\equref{cse7}, one needs to find suitable values for the bounds $\b_{rts}$. If $ITER$ is the number of iterations used in the Monte Carlo simulation algorithm for the estimation of $Q_{rts}$ then, one can consider, for example, the naive bound provided by the Central Limit Theorem with a $95\%$ confidence level
\begin{equation}\eqlabel{cse8}
\beta_{rts}=1.96\sqrt{\frac{\hat{Q}_{rts}(1-\hat{Q}_{rts})}{ITER}}.
\end{equation}
This bound has been used with some success for the two dimensional case (see \citet{HaimanPreda2}). As the authors pointed out, the main contribution to the total error is due to the simulation error, especially for small sizes of the window scan with respect to the scanning region. Our numerical study shows that Eq.\equref{cse8} is not feasible for the three dimensional case, the simulation error being to large with respect to the approximation error. Thus, for the simulation of $\hat{Q}_{rts}$, we use an importance sampling technique introduced in \citet{Naiman2001}. Next section illustrates how to adapt theirs algorithm to our problem.
\section{Simulation by importance sampling}\selabel{sec3}
\noindent
In this section we present a simulation method for $Q_{rts}$, which gives an unbiased estimate whose variance is typically smaller then that of the naive hit or miss Monte Carlo approach. The method is an adaptation of the importance sampling algorithm developed in \citet{Naiman2001} to our problem. The main idea behind is to express the tail of the scan distribution as a Bonferroni upper bound ($B$) with some correction factor ($\rho$). Let define for $1\leq i_{j}\leq N_{j}$, $j\in\{1,2,3\}$ the events $A_{i_1i_2i_3}=\{Y_{i_1i_2i_3}\geq\tau\}$. Then
\begin{align}\eqlabel{smeq1}
\PP\left(S_{m_1,m_2,m_3}(T_1,T_2,T_3)\geq\tau\right)&=\displaystyle\PP\left(\bigcup_{i_1=1}^{T_1-m_1+1}\bigcup_{i_2=1}^{T_2-m_2+1}\bigcup_{i_3=1}^{T_3-m_3+1}A_{i_1i_2i_3}\right)\nonumber\\
                                                    &=\displaystyle B\sum_{i_1=1}^{T_1-m_1+1}\sum_{i_2=1}^{T_2-m_2+1}\sum_{i_3=1}^{T_3-m_3+1}p_{i_1i_2i_3}I(i_1,i_2,i_3)\nonumber\\
                                                    &=B\rho,
\end{align}
where
\begin{equation}
\rho=\displaystyle \sum_{i_1=1}^{T_1-m_1+1}\sum_{i_2=1}^{T_2-m_2+1}\sum_{i_3=1}^{T_3-m_3+1}p_{i_1i_2i_3}I(i_1,i_2,i_3).
\end{equation}
Under the null hypothesis ($H_0$), $B$ is the Bonferroni upper bound given by
\begin{align}\eqlabel{smeq2}
B&=\displaystyle\sum_{i_1=1}^{T_1-m_1+1}\sum_{i_2=1}^{T_2-m_2+1}\sum_{i_3=1}^{T_3-m_3+1}{\PP(A_{i_1i_2i_3})}\nonumber\\
 &=(T_1-m_1+1)(T_2-m_2+1)(T_3-m_3+1)\PP(A_{111}),
\end{align}
$p_{i_1i_2i_3}$ defines an uniform probability distribution over $\{1,\dots,T_1-m_1+1\}\times\{1,\dots,T_2-m_2+1\}\times\{1,\dots,T_3-m_3+1\}$,
\begin{align}\eqlabel{smeq3}
p_{i_1i_2i_3}&=\frac{\PP(A_{i_1i_2i_3})}{\displaystyle \sum_{s_1=1}^{T_1-m_1+1}\sum_{s_2=1}^{T_2-m_2+1}\sum_{s_3=1}^{T_3-m_3+1}{\PP(A_{s_1s_2s_3})}}\nonumber\\
             &=\frac{1}{(T_1-m_1+1)(T_2-m_2+1)(T_3-m_3+1)},
\end{align}
and $I(i_1,i_2,i_3)=\displaystyle\int{\frac{1}{C(Y)}\frac{\mathbf 1_{A_{i_1i_2i_3}}}{\PP(A_{i_1i_2i_3})}d\PP}$ where $C(Y)$ represents the number of triples $(i_1,i_2,i_3)$ such that $Y_{i_1i_2i_3}$ exceeds the threshold $\tau$, that is
\begin{equation}\eqlabel{smeq4}
C(Y)=\displaystyle\sum_{i_1=1}^{T_1-m_1+1}\sum_{i_2=1}^{T_2-m_2+1}\sum_{i_3=1}^{T_3-m_3+1}{\mathbf 1_{A_{i_1i_2i_3}}}.
\end{equation}
Based on these identities the simulation algorithm (similar with the one described in \citet[page 303]{Naiman2001}) can be written as follows:
\begin{itemize}
\item[] Begin
    \begin{itemize}
    \item[] Repeat for each $k$ from $1$ to $ITER$ (iterations number)

        \begin{enumerate}
            \item[Step 1] Generate $T\in\{\tau,\dots\}$ according to the probabilities
            \begin{equation}
            p_T(t)=\frac{\PP(Y_{111}=t)}{\displaystyle\sum_{s\geq\tau}{\PP(Y_{111}=s)}},\ \ t\geq\tau\nonumber.
            \end{equation}
            \item[Step 2] Conditionally, given $T=t$, generate the triple $(J_1,J_2,J_3)$ in the set $\{1,\dots,T_1-m_1+1\}\times\{1,\dots,T_2-m_2+1\}\times\{1,\dots,T_3-m_3+1\}$ uniformly.
            \item[Step 3] Conditionally, given $T$ and $(J_1,J_2,J_3)$, generate the set of random variables $\{\tilde{Y}_{i_1i_2i_3}|J_s\leq i_s\leq J_s+m_s-1, s\in\{1,2,3\}\}$, uniformly from the set of all the vectors of length $m_1\times m_2\times m_3$ over the set of values taken by $Y_{i_1i_2i_3}$ and whose sum is equal with $T$. Take the remaining $\tilde{Y}_{i_1i_2i_3}$ to be i.i.d. and distributed according to the null hypothesis distribution.
            \item[Step 4] Take $c_k=C(\tilde{Y}_k)$ the number of all triples $(i_1,i_2,i_3)$ such that $\tilde{Y}_{i_1i_2i_3}\geq T$ and put $\hat{\rho}_k=\frac{1}{c_k}$.
        \end{enumerate}

    \item[] End Repeat
    \item[] Return $\hat{\rho}=\displaystyle\frac{1}{ITER}\sum_{k=1}^{ITER}{\hat{\rho}_k}$.
    \end{itemize}
\item[] End
\end{itemize}
\par\noindent
Clearly, $\hat{\rho}$ is an unbiased estimator for $\rho$ with estimated variance
\begin{equation}\eqlabel{smeq5}
Var(\hat{\rho})\approx \frac{1}{ITER-1}\displaystyle\sum_{k=1}^{ITER}{\left(\hat{\rho}_k-\frac{1}{ITER}\sum_{k=1}^{ITER}{\hat{\rho}_k}\right)^2}.
\end{equation}
For $ITER$ sufficiently large, as a consequence of CLT the error between the true and the estimated value of the tail $\PP\left(S_{m_1,m_2,m_3}(T_1,T_2,T_3)\geq\tau\right)$, corresponding to a $95\%$ confidence level, is given by
\begin{equation}\eqlabel{smeq6}
\beta=1.96B\sqrt{\frac{Var(\hat{\rho})}{ITER}}.
\end{equation}
Notice that for the simulation of $Q_{rts}$, we substitute $T_1$, $T_2$ and $T_3$ in the above relations with $r(m_1-1)$, $t(m_2-1)$ and $s(m_3-1)$ respectively. Therefore, we obtain the corresponding values for $\beta_{rts}$ as described by Eq.\equref{smeq6}.
\section{Numerical values for Binomial and Poisson models}\selabel{sec4}
\noindent
In this section, for selected values of the parameters of the binomial and Poisson distributions, we evaluate the approximation introduced in \seref{sec2} and provide the corresponding error bounds. We show the contributions of the approximation (Eq.\equref{cae8}) and simulation (Eq.\equref{cse7}) errors in the overall error.
\par\noindent
For all our simulations we used the importance sampling algorithm with $ITER=10^5$ replications. We compare our results with those existing in literature, see \citet{Glaz2010} for Bernoulli model, and with the simulated value of the scan statistics obtained by scanning the whole region $\MR$, denoted by $\hat{\PP}(S\leq n)$. The scanning of $\MR$ being more time consuming than the scanning of the subregions corresponding to $Q_{rst}$, we used $10^3$ repetitions of the algorithm.
\par\noindent
In Table~\ref{tab:Tb1}, we compare the results obtained by our approximation with the product type approximation presented by \citet{Glaz2010}. We observe that our approximation is very sharp.
\begin{table}[ht]
        \centering
        \caption{\scriptsize{Approximation for $\PP(S\leq n)$ in Bernoulli case: $m_1=m_2=m_3=5,T_1=T_2=T_3=60,ITER=10^5$}} 
        \label{tab:Tb1}
        \footnotesize{
            \begin{tabular}{c c c c c c c}
            \toprule
            $n$   & $\hat{\PP}(S\leq n)$ &     Glaz et al.              &      Our      & $E_{app}$             & $E_{sim}$                 & Total \\
                  &                      & Product type                 & Approximation &   Eq.\equref{cae8}    & Eq.\equref{cse7}          &  Error \\
            \midrule
                  &                      &                              &  $p=0.00005$  &                       &                           &        \\
            \cmidrule{4-4}\\
             $1$  & $0.841806$ & $0.841424$ & $0.851076$ & $0.011849$          & $0.064889$          & $0.076738$ \\
             $2$  & $0.999119$ & $0.999142$ & $0.999192$ & $0.000000$          & $0.000170$          & $0.000170$ \\
             $3$  & $0.999997$ & $0.999998$ & $0.999997$ & $0.000000$          & $3\times10^{-7}$    & $3\times10^{-7}$ \\
                  &                      &                              &              &                       &                           &        \\
                  &                      &                              &  $p=0.0001$  &                       &                           &        \\
            \cmidrule{4-4}\\
             $2$  & $0.993294$ & $0.993241$ & $0.993192$ & $0.000010$          & $0.001367$          & $0.001377$ \\
             $3$  & $0.999963$ & $0.999964$ & $0.999963$ & $0.000000$          & $0.000005$          & $0.000005$ \\
             $4$  & $0.999999$ & $0.999999$ & $0.999999$ & $0.000000$          & $2\times10^{-9}$    & $2\times10^{-9}$ \\
            \bottomrule
            \end{tabular}}
\end{table}
\par\noindent
Table~\ref{tab:Tb2} presents the numerical results obtained by scanning the region $\MR$ of size $60\times 60\times 60$ with two windows of the same volume but different sizes, first a cubic window of size $4\times 4\times 4$ and second a rectangular region of size $8\times 4\times 2$. We observe that the results are closely related, but significantly different.
\begin{table}[ht]
        \centering
        \caption{\scriptsize{Approximation for $\PP(S\leq n)$ over the region $\MR$ with windows of the same volume by different sizes: $T_1=T_2=T_3=60,p=0.0025,ITER=10^5$}} 
        \label{tab:Tb2}
        \footnotesize{
            \begin{tabular}{c c c c c c}
            \toprule
            $n$   & $\hat{\PP}(S\leq n)$ &       Our      & $E_{app}$             & $E_{sim}$                 & Total \\
                  &                      &  Approximation &   Eq.\equref{cae8}    & Eq.\equref{cse7}          &  Error \\
            \midrule
            \multicolumn{6}{c}{$m_1=m_2=m_3=4$}\\
            \cmidrule{3-5}\\
             $5$  & $0.961691$ &  $0.963506$ & $0.000038$          & $0.003622$          & $0.003660$ \\
             $6$  & $0.999006$ &  $0.999023$ & $0.000000$          & $0.000071$          & $0.000071$ \\
             $7$  & $0.999980$ &  $0.999980$ & $0.000000$          & $0.000001$          & $0.000001$  \\
             $8$  & $0.999999$ &  $0.999999$ & $0.000000$          & $2\times10^{-9}$    & $2\times10^{-9}$ \\
                  &                      &                              &   &                       &          \\
            \multicolumn{6}{c}{$m_1=8,m_2=4,m_3=2$}\\
            \cmidrule{3-5}\\
             $5$  & $0.969189$ &  $0.969110$ & $0.000007$          & $0.003387$          & $0.003395$ \\
             $6$  & $0.999297$ &  $0.999228$ & $0.000000$          & $0.000071$          & $0.000071$ \\
             $7$  & $0.999984$ &  $0.999984$ & $0.000000$          & $0.000001$          & $0.000001$ \\
             $8$  & $0.999999$ &  $0.999999$ & $0.000000$          & $2\times10^{-9}$    & $2\times10^{-9}$ \\
            \bottomrule
            \end{tabular}}
\end{table}
\par\noindent
In Table~\ref{tab:Tb3} we have included numerical values emphasizing the situation described by \reref{rem1}. We consider the Bernoulli model of parameter $p=0.0001$ over the region $\MR$ of size $185\times 185\times 185$ and scan it with a cubic window of length $10$. The second and forth columns gives the values corresponding to the bounds described in Eq.\equref{meq14}, while in the third column we presented the simulated values for $\PP\left(S_{10,10,10}(185,185,185)\leq n\right)$.
\begin{table}[ht]
        \centering
        \caption{\scriptsize{Approximation for $\PP(S\leq n)$ based on Eq.\equref{meq14}: $m_1=m_2=m_3=10,T_1=T_2=T_3=185,L_1=L_2=L_3=20,ITER=10^5$}} 
        \label{tab:Tb3}
        \footnotesize{
            \begin{tabular}{c c c c}
            \toprule
            $n$   & $\PP\left(S(L_1+1,L_2+1,L_3+1)\leq n\right)$ &      $\hat{\PP}(S\leq n)$      & $\PP\left(S(L_1,L_2,L_3)\leq n\right)$ \\
            \midrule
             $4$  & $0.97524633$      &  $0.97465263$       & $0.97491935$          \\
                  & \tiny{$(\pm0.00754004)$}   &  \tiny{$(\pm0.00618987)$}               & \tiny{$(\pm0.00643099)$}     \\
             $5$  & $0.99931055$      &  $0.99935163$       & $0.99938629$          \\
                  & \tiny{$(\pm0.00015833)$}   &  \tiny{$(\pm0.00014759)$}               & \tiny{$(\pm0.00013490)$}    \\
             $6$  & $0.99998641$      &  $0.99998632$       & $0.99998784$          \\
                  & \tiny{$(\pm0.00000272)$}   &  \tiny{$(\pm0.00000326)$}               & \tiny{$(\pm0.00000230)$}    \\
            \bottomrule
            \end{tabular}}
\end{table}
\par\noindent
In order to compare the binomial and Poisson models, in Table~\ref{tab:Tb4}, we have evaluated the distribution of the scan statistics over a region of size $84\times 84\times 84$ scanned with a $4\times 4\times 4$ cubic window, in the two situations. In the first case we have a binomial random field with parameters $m$ and $p$, that is $X_{ijk}\sim B(m,p)$, while in the second we considered that $X_{ijk}\sim P(\lambda)$, with $\lambda=mp$.
\begin{table}[ht]
        \centering
        \caption{\scriptsize{Approximation for $\PP(S\leq n)$ in Binomial and Poisson cases: $m_1=m_2=m_3=4,T_1=T_2=T_3=84,ITER=10^5$}} 
        \label{tab:Tb4}
        \footnotesize{
            \begin{tabular}{c c c c c c}
            \toprule
            $n$   & $\hat{\PP}(S\leq n)$ &       Our      & $E_{app}$             & $E_{sim}$                 & Total \\
                  &                      &  Approximation &   Eq.\equref{cae8}    & Eq.\equref{cse7}          &  Error \\
            \midrule
                  &                      &    $Binomial:$ &  $m=10,p=0.0025$      &                       &          \\
            \cmidrule{3-4}\\
             $10$  & $0.726386$ &  $0.723224$ & $0.007763$          & $0.032197$          & $0.039960$ \\
             $11$  & $0.954605$ &  $0.955417$ & $0.000123$          & $0.003079$          & $0.003202$ \\
             $12$  & $0.993938$ &  $0.993906$ & $0.000001$          & $0.000331$          & $0.000333$  \\
             $13$  & $0.999289$ &  $0.999284$ & $0.000000$          & $0.000033$          & $0.000033$ \\
             $14$  & $0.999923$ &  $0.999921$ & $0.000000$          & $0.000003$          & $0.000003$  \\
             $15$  & $0.999992$ &  $0.999992$ & $0.000000$          & $3\times10^{-7}$    & $3\times10^{-7}$  \\
                  &                      &                              &   &                       &          \\
                  &                      &    $Poisson:$  &  $\lambda=0.025$      &                       &          \\
            \cmidrule{3-4}\\
             $10$  & $0.713184$ &  $0.708481$ & $0.009211$          & $0.035294$          & $0.044506$ \\
             $11$  & $0.950947$ &  $0.950197$ & $0.000143$          & $0.003345$          & $0.003488$ \\
             $12$  & $0.993624$ &  $0.993452$ & $0.000002$          & $0.000365$          & $0.000367$ \\
             $13$  & $0.999218$ &  $0.999210$ & $0.000000$          & $0.000038$          & $0.000038$ \\
             $14$  & $0.999912$ &  $0.999911$ & $0.000000$          & $0.000003$          & $0.000003$  \\
             $15$  & $0.999990$ &  $0.999990$ & $0.000000$          & $3\times10^{-7}$    & $3\times10^{-7}$  \\
            \bottomrule
            \end{tabular}}
\end{table}
\par\noindent
Notice that the contribution of the approximation error ($E_{app}$) to the total error is almost negligible in most of the cases with respect to the simulation error ($E_{sim}$). Thus, the precision of the method will depend mostly on the number of iterations ($ITER$) used to estimate $Q_{rts}$.
\par\noindent
The time required for the computations presented in this section was about two hours for each table on a computer of medium performances. The programs are written in MATLAB and are available from the authors.
\section{Conclusions}\selabel{sec5}
\noindent
In this article we derived an approximation for the three dimensional discrete scan statistic viewed as the maximum of a $1$-dependent stationary sequence of random variables. We also provide the corresponding theoretical and simulation error bounds. In the three dimensional scan statistics framework, it is essential to reduce the variance of simulated values. For this purpose we used an importance sampling method. A simulation study for the binomial and Poisson models shows the accuracy as well as the limit of our method.

\end{document}